\documentclass[12pt, epsfig]{article}
\usepackage[dvips]{color}
\usepackage{epsfig}%
\usepackage[active]{srcltx}%
\usepackage{amsmath,amsfonts,amssymb,amsthm,amstext,amscd,eucal,srcltx}
\usepackage{epsfig,graphicx,bm}
\usepackage{epstopdf, epsf}
\usepackage{dcolumn}
\usepackage{hyperref}
\usepackage{subfigure}
\usepackage{textcomp}

\definecolor{darkgreen}{rgb}{0,0.3,0}
\definecolor{darkblue}{rgb}{0,0,0.3}
\definecolor{darkred}{rgb}{0.7,0,0}

\newcommand{\be}{\begin{equation}}
\newcommand{\bse}{\begin{subequations}}
\newcommand{\ese}{\end{subequations}}
\newcommand{\bea}{\begin{eqnarray}}
\newcommand{\eea}{\end{eqnarray}}
\newcommand{\ba}{\begin{array}}
\newcommand{\ea}{\end{array}}
\newcommand{\ee}{\end{equation}}

\makeatletter \@addtoreset{equation}{section}

\def\dre_g{\delta\rho_g}
\def\dpe_g{\delta P_g}
\def\dqe_g{\delta q_g}
\def\dre{\delta\rho}
\def\dpe{\delta P}
\def\dqe{\delta q}

\def\YM1{\frac{\dot\phi^2}{a^2}}
\def\YM2{\frac{g^2\phi^4}{a^4}}

\begin{document}
\title {Gauge inflation by kinetic coupled gravity}
\vspace{3mm}

\author{F. Darabi\thanks{Email: f.darabi@azaruniv.edu}, A. Parsiya\thanks{Email: a.parsiya@azaruniv.edu}\\
{\small Department of Physics, Azarbaijan Shahid Madani University, Tabriz 53714-161, Iran.}}

\maketitle

\begin{abstract}
Recently, a new class of inflationary models, so called \emph{gauge-flation} or non-Abelian gauge field inflation has been introduced where the slow-roll inflation is driven by a non-Abelian gauge field $\textbf{A}$ with the field strength $\textbf{F}$. This class of models are based on a gauge field theory having $\textbf{F}^2$ and $\textbf{F}^4$ terms with a non-Abelian gauge group minimally coupled to gravity. Here, we present a new class of such inflationary models based on a gauge field theory having only $\textbf{F}^2$ term with non-Abelian gauge fields non-minimally coupled to gravity. The non-minimal coupling is set up by introducing the Einstein tensor besides the metric tensor within the $\textbf{F}^2$ term, which is called kinetic coupled gravity. A perturbation analysis is performed to confront the inflation under consideration with Planck and BICEP2 results.\\
\\
Keywords: Gauge-flation, Non-Abelian gauge field, Kinetic coupled gravity.
\\
PACS: {98.80.Cq, 98.80.-k, 98.80.Es.}
\end{abstract}

\maketitle

\section{Introduction}

The idea of inflation is a well known scenario to overcome the problems of
standard cosmology \cite{Inflation-Books, Weinberg}. It tells us that the early universe has experienced an inflationary expansion phase in a very short period of time. This scenario is also very successful in reproducing the current cosmological data through the $\Lambda$CDM model \cite{Bassett-review}. One of the key ingredients in inflation is the existence of a {\it scalar field} which is subjected to the {\it slow-roll} approximation, where the kinetic energy of the scalar field remains sufficiently small compared to its potential energy. The universe experiences an inflationary expansion
in the slow-roll regime, and once the kinetic energy becomes comparable to its potential energy the inflation is ended. The scalar field gets more and
more kinetic energy and reaches the minimum of potential, afterwards the scalar field starts fast frictional oscillations around this minimum and transfer its kinetic energy into the matter or radiation to establish the reheating phase. Inflationary models typically benefit of single or multi-scalar field theories. However, it is very appealing to model the inflationary scenario within the particle physics context. This is because the inflation is supposed
to happen at very early universe where use of high energy physics is inevitable. Other than scalar fields, the vector and vector gauge fields are among the most suitable candidates to set up inflation in the context of high energy physics \cite{Higgs-inflation}-\cite{vector-inflation2}. Nevertheless, a successful vector inflation is not possible in a \emph{gauge invariant} manner \cite{vector-inflation-loophole}.

To overcome this problem, a new class of vector inflation models is proposed in the framework of gauge field theories, so called {\it Gauge-flation }\cite{Maleknejad}. To remedy the incompatibility between vector gauge fields in high energy physics and requirement of isotropy in cosmology, three gauge fields are introduced with rotational symmetry resulting from $SU(2)$ non-Abelian gauge transformations. The global part of this gauge symmetry is then identified with the rotational symmetry in 3$d$ space and so the isotropy of space in the presence of this vector gauge field is preserved. 

Recently, in order to justify the current acceleration of universe, a considerable amount of activity has been focused on the kinetic coupled gravity theories
\cite{Sushkov:2009, Gao:2010, Germani, Nojiri}. In these theories, a non-canonical kinetic term or a non-minimal coupling to the metric is responsible for the physics of acceleration. In a very recent work, the authors have introduced a new scenario of scalar field inflation in the context of kinetic coupled gravity \cite{Darabi}. In the present work, we aim to generalize the scenario of scalar field inflation to gauge field inflation in the context of kinetic coupled gravity. In this regard, we combine the scenario of gauge-flation \cite{Maleknejad} with the idea of kinetic coupled gravity. In explicit words, we replace $\textbf{F}^2$ and $\textbf{F}^4$ terms minimally coupled to gravity by $\textbf{F}^2$ term non-minimally coupled to gravity. The non-minimal coupling is established by considering the Einstein tensor besides the metric tensor within the $\textbf{F}^2$ term, which is called kinetic coupled gravity. 

\section{\textbf{Gauge Inflation by non-Abelian $F^2$ and $F^4$ terms minimally coupled to gravity}}

In this section we briefly introduce the idea of gauge inflation extensively
discussed in \cite{Maleknejad}. Let us consider a 4-dimensional $SU(2)$ gauge field $A^a_{~\mu}$, where
$a,b, ...$ and $\mu,\nu, ...$ are used respectively for the indices
of the gauge algebra and the space-time coordinates. Gauge and Lorentz invariant Lagrangians ${\cal L}(F^a_{~\mu\nu}, g_{\mu\nu})$ are constructed out of the metric tensor $g_{\mu\nu}$ and the field strength tensor 
\be\label{F-general1}%
F^a_{~\mu\nu}=\partial_\mu A^a_{\nu}-\partial_{\nu}
A^a_{\mu}-g\epsilon^a_{bc}A^b_{\mu}A^c_{\nu}, %
\ee%
where $\epsilon_{abc}$ is the totally antisymmetric tensor. We take flat
($k=0$) FRW background metric%
\be \label{FRW} ds^2=-dt^2+a(t)^2\delta_{ij}dx^idx^j, %
\ee%
where indices $i,j, ...$ label the  spatial coordinates. We know that the
standard problem of anisotropy is a typical problem for the vector models of inflation. Hence, in the first place one should solve this problem in
the context of present gauge-vector model of inflation.
In this regard, we explain how the effective inflaton field is introduced in the present model. 
If we consider an Abelian gauge field and turn on a vector gauge field in the background, then the isotropy  or rotational symmetry of the flat FRW geometry will
be broken down, unless only we turn on the temporal component which should be only time dependent in order to preserve homogeneity. Hence, this gauge field becomes a pure gauge with vanishing field strength which is not physically
viable. However, one may consider non-Abelian gauge fields with gauge group indices. It is obvious that two gauge fields which are related by a local gauge transformation are physically equivalent.
In the coordinate system of standard spatially flat FRW metric subject to
the homogeneity we may turn on a typical vector gauge field with spatial components $A_i=A_i(t)$. The time component $A_0$ may always be set to zero in the temporal gauge. This choice fixes the gauge transformations
up to time independent gauge transformations. On the other hand, since we have already ignored the space dependence of the gauge fields, it turns out
that the gauge freedom is fixed up to space-time independent global gauge transformations. Therefore, up to global gauge transformations, the background gauge field is given by
\begin{equation}
A_0=0, \,\,\,\, A_i=A_i(t).
\end{equation}
The global gauge transformations can remedy the problem of anisotropy emerged
by non-Abelian gauge fields. Two non-Abelian gauge fields related by a gauge transformation are physically equivalent as follows
\begin{equation}
(A_i)_G=U^{-1}A_i U.
\end{equation}
Moreover, the global spatial rotations are defined by
\begin{equation}
(A_i)_R=R_{ij}A_j.
\end{equation}
Now, if one chooses $(A_i)_R=(A_i)_G$, then the anisotropy or rotational non-invariance caused by turning on vector fields in the background may be compensated by the global gauge transformation, and the physical gauge configuration may preserve the isotropy or rotational symmetry \cite{Maleknejad}.  

Bearing this in mind, we work in the temporal gauge $A^a_{0}=0$, and in order to respect for the cosmological principle at the background level, namely homogeneity, we just allow for $t$ dependent field configurations
\be\label{A-ansatz-background}
A^a_{~\mu}=\left\{
\begin{array}{ll} \phi(t)\delta^a_i\, ,\qquad  &\mu=i
\\
0\,\,\,\,\,\,\,\,\,\,\,\,\,\,, \qquad &\mu=0\,.
\end{array}\right.
\ee%
By identifying the gauge indices with the spatial indices, the rotation group $SU(2)$ is identified with the global part of the gauge group, as discussed
above. Note that $\phi(t)$ by itself is not a scalar under the general coordinate transformations, while 
\be\label{psi-def} 
\psi(t)=\frac{\phi(t)}{a(t)},%
\ee%
is indeed a scalar. Using \eqref{F-general1}, the components of the field strengths in this ansatz are obtained
\be\label{F-components}%
F^a_{~0i}=\dot{\phi}\delta^a_i\,,\qquad
F^a_{~ij}=-g\phi^2\epsilon^a_{~ij}.
\ee%
By fixing the gauge as $A^a_{~0}=0$, the system is left with nine degrees of freedom, $A^a_{~i}$. The trivial choice for the action is the Yang-Mills action minimally coupled to Einstein gravity. It is known that in any cosmological model subject to a matter field, the necessary condition for a successful inflation is $\rho+3p<0$ where $\rho$ and $p$ are the energy density and pressure of the matter field, respectively. It is easy to see that the Yang-Mills action minimally coupled to Einstein gravity will not lead to the inflation. This is because as a result of scale invariance of the Yang-Mills action one immediately obtains $p=\rho/3$ and $\rho\geq 0$, which obviously do not satisfy the above necessary condition. Hence, we need to modify the Yang-Mills action minimally coupled to Einstein gravity. One such appropriate choice has recently been considered in \cite{Maleknejad}
\begin{equation}\label{The-model}%
S=\int
d^4x\sqrt{-{g}}\left[-\frac{R}{2}-\frac{1}{4}F^a_{~\mu\nu}F_a^{~\mu\nu}+\frac{\kappa}{384}
(\kappa^{\mu\nu\lambda\sigma}F^a_{~\mu\nu}F^a_{~\lambda\sigma})^2\right],
\end{equation}
which las led to a successful inflation. The appearance of $F^4$ term
in this action has been justified by the requirement that the contribution of this term to the energy momentum tensor results in the equation of state
$p =-\rho$, which makes it suitable for driving inflationary dynamics. However,
a reasonable justification for this term needs rigorous quantum gauge field theory analysis and some particle physics settings \cite{Maleknejad}.

\section{\textbf{Gauge Inflation by non-Abelian $F^2$ term non-minimally coupled to gravity }}

In this section, we introduce a new action where the Yang-Mills
field is non-minimally coupled to Einstein gravity as follows
\begin{equation}\label{The-model1}%
S=\int
d^4x\sqrt{-{g}}\left[-\frac{R}{2}-\frac{1}{4}(g^{\rho\mu}+\alpha G^{\rho\mu})( g^{\nu\lambda}+\alpha G^{\nu\lambda})F_{~a\rho\lambda}F^a_{~\mu\nu}\right]\,,
\end{equation}
where $g^{\mu\nu}$ is the metric tensor, $G^{\mu\nu}$ is the Einstein tensor and $\alpha>0$ are constant parameters. In this action, we just
use one $F^2$ term instead of $F^2$ and $F^4$ terms considered in the previous action \cite{Maleknejad}. The price we pay for this replacement is to introduce the Einstein tensor besides the metric tensor as the geometric coupling of two $F$ terms, namely we work with $F^2$ term non-minimally coupled to gravity. We aim to show that it is possible to have inflation in this Yang-Mills field non-minimally coupled to gravity. In this direction, we first obtain the energy momentum tensor \begin{eqnarray}
T_{\alpha\beta}&=&\{F_a{^\mu_\alpha}F^a_{\mu\beta}-\frac{1}{4}g_{\alpha\beta}F_{a\mu\nu}F^{a\mu\nu}\}
\nonumber\\
&+&\frac{\alpha\kappa}{4}\{g_{\alpha\beta}g^{\rho\mu}G^{\nu\lambda}F_{a\rho\lambda}F{^a_{\mu\nu}}
-2G^{\rho\mu}F_{a\rho\beta}F^a_{\mu\alpha}+\tilde{g}_{\alpha\beta}R
+\tilde{g}_{\lambda\nu}\tilde{g}^{\lambda\nu}R_{\alpha\beta}
-4\tilde{g}_{\lambda\beta}R^\lambda_\alpha\},
\end{eqnarray}
where $\tilde{g}_{\lambda\nu}=g^{\rho\mu}F_{a\rho\lambda}F^a_{\mu\nu}$.
We have retained just the linear terms in $\alpha$, for simplicity. Actually, the linear term $\alpha G^{\mu\nu}$ contributing to the kinetic term in the present model of inflation effectively plays the role of inflaton potential in the ordinary models of inflation. Hence, similar to the ordinary models of inflation having large inflaton potential, the order of magnitude of $\alpha G^{\mu\nu}$ in the kinetic term should be considerably larger than that of $g^{\mu\nu}$, at early universe. Therefore, it is trivial that the terms quadratic in $\alpha$ should also be considered in the energy-momentum tensor. However, for simplicity and to avoid any complicate nonlinear equations, we have retained the terms linear in $\alpha$. This simplification may be justified because if we can succeed to provide inflation using the linear terms $\alpha G^{\rho\mu}F_{\rho}^{a\nu}F^{a}_{\mu\nu}$, playing the role of large inflaton potential, we can succeed as well to provide inflation using the quadratic positive definite terms $\alpha^2 G^{\rho\mu}G^{\nu\lambda}F^{a}_{\rho\lambda}F^{a}_{\mu\nu}$, because these terms will certainly strengthen the role of kinetic coupled Einstein tensor as a large inflaton potential.

The action \eqref{The-model1} differs from the non-minimal theories introduced in \cite{Bamba}, where non-abelian and non-minimal vector-$F(R)$ gravity theories are studied to provide inflation. The main differences lie in the presence of $F(R)$ gravity and also potential terms for Yang-Mills
field in these actions. Noting that the actions in \cite{Bamba} have the advantage of being in the class of modified gravity theories, one may point out that the model described by the action \eqref{The-model1} has the advantage that it is economic because of using a free Yang-Mills field rather than a Yang-Mills field with a potential. In explicit words, it is more reasonable to think that the universe prefers an economic action to trigger the inflation by the least elements: geometry and a free Yang-Mills field.

Using the Friedmann-Robertson-Walker (FRW) background \eqref{FRW} and the gauge field ansatz \eqref{A-ansatz-background}, we may cast the above energy-momentum
tensor in the form of a homogenous perfect fluid where
\begin{eqnarray}\label{density}
T_{00}=\rho=\frac{3}{2}(\frac{\dot{\phi}^2}{a^2}+\frac{g^2\phi^4}{a^4})+\frac{\alpha}{4}
\{3(\frac{\dot{\phi}^2}{a^2}+2\frac{g^2\phi^4}{a^4})({2\dot{H}+3H^2})
\nonumber\\
-9H^2(\frac{\dot{\phi}^2}{a^2})
+3(\frac{\dot{\phi}^2}{a^2})R-6(\frac{\dot{\phi}^2}{a^2}-\frac{g^2\phi^4}{a^4})R_{00}\},
\end{eqnarray}
\begin{eqnarray}\label{P}
T_{ii}=p=\frac{1}{2}(\frac{\dot{\phi}^2}{a^2}+\frac{g^2\phi^4}{a^4})+\frac{\alpha}{4}
\{3(\frac{\dot{\phi}^2}{a^2}+2\frac{g^2\phi^4}{a^4})({2\dot{H}+3H^2})-9H^2(\frac{\dot{\phi}^2}{a^2})
\nonumber\\
+[-3R+6R^i_i](\frac{\dot{\phi}^2}{a^2})+[6R-18R^i_i](\frac{g^2\phi^4}{a^4})\},
\end{eqnarray}
and
\be\label{R} %
R= 6[\dot{H}+2H^2] ,\qquad R^0_0=3[\dot{H}+H^2],\qquad R^i_i=[\dot{H}+3H^2]. %
\ee%
We may divide $\rho$ and $p$ in the following way
\be\label{rho-P-total} %
\rho= \rho_{_{YM}}+\rho_\alpha\ ,\qquad p=p_{_{YM}}+p_\alpha, %
\ee%
where
\be\label{R1} %
\rho_{_{YM}}=\frac{3}{2}(\frac{\dot{\phi}^2}{a^2}+\frac{g^2\phi^4}{a^4}),\qquad p_{_{YM}}=\frac{1}{2}(\frac{\dot{\phi}^2}{a^2}+\frac{g^2\phi^4}{a^4}), %
\ee%
\be\label{R2} %
\rho_{\alpha}=\frac{3\alpha}{2}
[({\dot{H}+3H^2})(\frac{\dot{\phi}^2}{a^2})-{\dot{H}}(\frac{g^2\phi^4}{a^4})], %
\ee%
\be\label{R3} %
p_{\alpha}=-\frac{3\alpha}{2}
[({\dot{H}+3H^2})(\frac{\dot{\phi}^2}{a^2})-({5\dot{H}+6H^2})(\frac{g^2\phi^4}{a^4})]. %
\ee%
Substituting (\ref{R}) into  (\ref{R2}) and (\ref{R3}), we obtain
\be\label{R2'} %
\rho_{\alpha}=\frac{3\alpha}{2}
[(\dot{H}+3H^2)x-\dot{H}y], %
\ee%
\be\label{R3'} %
p_{\alpha}=-\frac{3\alpha}{2}
[(\dot{H}+3H^2)x-(5\dot{H}+6H^2)y], %
\ee%
where
\be\label{xy} %
x=\frac{\dot{\phi}^2}{a^2}\,\,\, , \qquad y=\frac{g^2\phi^4}{a^4}. %
\ee%
Now, bearing in mind that in general the inflation needs a negative pressure,
and noting that $p_{_{YM}}=\frac{1}{3}\rho_{_{YM}}$, we require that the effective pressure corresponding to the Einstein tensor in the kinetic term be negative, by demanding the vacuum equation of state $p_\alpha=-\rho_\alpha$. This gives the scalar field equa  
\be\label{R4} %
\frac{d^2y}{dt^2}-2(2\dot{H}+3H^2)y=0, %
\ee%
which relates the dynamics of scalar field to that of scale factor. Using the condition $p_\alpha=-\rho_\alpha$ in the Einstein equations and also \eqref{R}, we obtain the corresponding Friedmann equations
\begin{eqnarray}\label{Friedmann1}
3H^2=\left( \frac{3}{2}+\frac{3\alpha}{2}A\right)x
+\left( \frac{3}{2}-\frac{3\alpha}{2}B\right)y,
\end{eqnarray}
\begin{eqnarray}\label{Friedmann2}
\dot{H}=-(x+y),
\end{eqnarray}
where 
\bse
\begin{eqnarray}\label{AB}
A&=&({\dot{H}+3H^2}), \\  B&=&{\dot{H}}.
\end{eqnarray}
\ese
Now, in order to work out with the slow-roll approximation we define the slow-roll parameters
\be\label{epsilon-eta-rho-P}
\varepsilon\equiv -\frac{\dot H}{H^2}=\frac32\frac{\rho+p}{\rho}\,,\qquad \eta=\varepsilon-\frac{\dot\varepsilon}{2H\varepsilon}\,,
\ee
from which one may specify the slow-roll dynamics by $\varepsilon,\ \eta\ll 1$. Using the Friedmann equations
\begin{eqnarray}\label{friedmann0}
3H^2&=\rho=(\rho_{_{YM}}+\rho_\alpha), \\\nonumber  2\dot{H}+3H^2&=-p=-(p_{_{YM}}+p_\alpha)\,,
\end{eqnarray}
where $p_\alpha=-\rho_\alpha$ and $p_{_{YM}}=\dfrac{1}{3}\rho_{_{YM}}$, we obtain%
\be\label{epsilon-rho0-rho1}
\varepsilon= \frac{2\rho_{_{YM}}}{\rho_{_{YM}}+\rho_\alpha}\,.
\ee%
Therefore, to have the slow-roll approximation the $\alpha$-term contribution
$\rho_\alpha$ should dominate over the Yang-Mills contribution
$\rho_{_{YM}}$, namely $\rho_\alpha\gg \rho_{_{YM}}$. In fact, considering Eq.\eqref{R2} and Eq.\eqref{R3}, we realize that although $\alpha$ is small, but at the beginning of inflation we need a huge $H$ to drive the
inflation, hence the first term of $\rho_\alpha$ in \eqref{R2} containing $\alpha({2\dot{H}+3H^2})$ can dominate not only over the second term, but also dominates over $\rho_{_{YM}}$ defined in Eq.\eqref{R1}. 

However, note that it is not enough to make sure $\varepsilon\ll 1$ for a successful inflation. In fact, time-variations of $\varepsilon$ and all the other physical dynamical variables of the problem, like $\eta$ and the $\psi$ field, must also remain small during the inflation over a reasonable period of time. For typical slow-roll models, $\eta$ usually measures the rolling velocity of the inflaton, namely $\eta=-\frac{\ddot{\phi}}{H\dot{\phi}}$. Hence, it is useful to define \cite{Maleknejad}%
\be\label{delta-def}
\delta\equiv-\frac{\dot{\psi}}{H\psi}\,,
\ee%
which is related to $\varepsilon$ and $\eta$ through the equations 
\bse
\begin{eqnarray}
\label{epsil}
\varepsilon&=&2-{\alpha}\left\lbrace (1-\delta)^2 A-B \gamma \right\rbrace\psi^2 , \\ \label{tilde-eta}
\eta&=&\varepsilon-{(2-\varepsilon)}\left(-\frac{\delta}{\varepsilon}+\frac{\dot{Z}}{2HZ\varepsilon}\right)\,,
\end{eqnarray}
\ese
and 
\bse
\begin{eqnarray}
\label{Z}
Z&=A-\gamma B, \\  \gamma &=\dfrac{g^2\psi^2}{H^2}\label{gamma}.
\end{eqnarray}
\ese
The slow-roll regime requires $\dot\varepsilon\sim H\varepsilon^2$ and $\eta\sim\varepsilon$, so we demand in \eqref{tilde-eta} that 
\bse
\begin{eqnarray} 
\label{Z1}
\delta\simeq \varepsilon^2,\\ \label{Z11}
\frac{\dot{Z}}{2HZ}\simeq\varepsilon^2.
\end{eqnarray}
\ese
Now, using  and \eqref{Z} in \eqref{epsil}
we have
\begin{eqnarray}
\label{epsi10}
\varepsilon&=&2-{\alpha}Z\psi^2. 
\end{eqnarray}
Using Eqs.\eqref{xy}, \eqref{Friedmann1}, \eqref{friedmann0}, \eqref{delta-def}
, \eqref{gamma} and $\delta\approx0$ we obtain
\begin{eqnarray}\label{rho2}
\rho=\left( \frac{3}{2}+\frac{3\alpha}{2}A\right)H^2\psi^2+\left( \frac{3}{2}-\frac{3\alpha}{2}B\right)\gamma H^2\psi^2,
\end{eqnarray}
or
\begin{eqnarray}\label{rho3}
\rho\approx H^2\psi^2\left[\frac{3}{2}(1+\gamma)+\frac{3\alpha}{2}Z \right].
\end{eqnarray}
Explicitly, the equations \eqref{friedmann0}, \eqref{tilde-eta}, \eqref{epsi10}, and \eqref{rho3} admit the solutions %
\bse 
\begin{eqnarray} \label{epsilon-x}
\varepsilon&\simeq\psi^2(\gamma+1),\\
\label{delta-x}
\delta&=\frac{\varepsilon}{2}(\varepsilon-\eta)+\frac{\dot{Z}}{2HZ},%
\end{eqnarray}\ese%
where $\simeq$ means equality to first order in the slow-roll parameter $\varepsilon$ and 
\be\label{x-def}
\gamma= \frac{g^2\psi^2}{H^2}\,,\qquad \Longleftrightarrow\qquad
H^2\simeq\frac{g^2\varepsilon}{\gamma(\gamma+1)}\,.%
\ee%
Using $p_\alpha=-\rho_\alpha$, the conservation equation for the Yang-Mills
component of the perfect fluid becomes
\begin{equation}\label{conservation}
{\dot{\rho}_{_{YM}}+3H(\rho_{_{YM}}+p_{_{YM}})\neq 0}.
\end{equation}
One realizes that the main reason of energy non-conservation is the
energy transfer to the scalar field whose origin is nothing but the Einstein tensor contributing to the kinetic energy of the scalar field. In other
words, the Einstein tensor which indicates the curvature
of spacetime at early universe, with huge order of magnitude $H\gg 1$
, induces an effective energy-momentum
tensor ($\rho_{\alpha}, p_{\alpha}$) which contributes to the kinetic energy of the
scalar field.
From \eqref{x-def} and using  $\delta\approx\varepsilon^2$ and $H=constant$
we obtain
\begin{equation}\label{dotgamma=0}
{\dot{\gamma}\simeq0},
\end{equation}
which shows that $\gamma$ is a positive parameter which is slowly varying during the slow-roll regime. Considering \eqref{gamma} with $H\approx {const}$
and that $\gamma$ is a slowly varying parameter, we find that $\psi$ is slowly varying, too. Hence, \eqref{gamma} implies that $\gamma H^2$ is almost constant during the slow-roll regime from which we obtain \cite{Maleknejad}
\be\label{epsilon-H}%
\frac{\gamma}{\gamma_i}\simeq\frac{H_i^2}{H^2}\,\Longleftrightarrow
\frac{\varepsilon}{\varepsilon_i}\simeq\frac{\gamma+1}{\gamma_i+1}\,,
\ee%
where $\varepsilon_i,\ \gamma_i$ and $H_i$ are the values of parameters at
the beginning of inflation. It is obvious that the slow-roll regime ends when $\varepsilon_f\approx1$. This happens when the scale factor becomes
inflationary large and $H$ becomes small towards the end of inflation and the contribution of $\alpha G^{\mu\nu}$ in the kinetic term of the gauge fields starts decreasing towards strengthening the condition $\rho_\alpha\sim \rho_{_{YM}}$ which runs $\varepsilon_f$ towards 1. Therefore, we find \cite{Maleknejad}
\be 
\gamma_f\simeq \frac{\gamma_i+1}{\varepsilon_i} \,,\qquad
\frac{H_f^2}{H_i^2}\simeq \frac{\gamma_i}{\gamma_i+1}\ \varepsilon_i\,.
\ee
Now, we can compute the number of e-folding \cite{Maleknejad}
\bea\label{Ne-gauge-flation}
N_{e}=\int_{t_i}^{t_f} Hdt=-\int_{H_i}^{H_f} \frac{dH}{\varepsilon
H}\simeq \frac{\gamma_i+1}{2\varepsilon_i}
\ln\frac{\gamma_i+1}{\gamma_i}\,,
\eea
where use has been made of $\varepsilon\equiv -\frac{\dot H}{H^2}$. It is
seen that number of e-folding is inversely related to the slow roll parameter
$\varepsilon_i$, so the more small slow roll parameter, the more e-folding.

\section{Perturbation analysis}

In this section, we study the perturbation theory at the smallest level over the spatially flat FRW isotropic background by considering the perturbation of inflaton field. Scalar and tensor perturbations emerging during the inflation might have left an imprint in the CMB anisotropy and on the LSS \cite{4,5}. Usually, for scalar and tensor fluctuations the power spectrum and spectral index are given respectively by $\Delta_R^2(k)$, $n_R$ and $\Delta_T^2(k)$, $n_T$. The curvature perturbation power spectrum is given by \cite{Lyth}
\begin{eqnarray}\label{16}
\Delta_R^2(k)=\left[\left(\frac{H}{\dot{\psi}}\right)^2\langle|\delta\psi|^2\rangle\right]|_{k=aH},
\end{eqnarray}
where $k$ is the co-moving wavenumber. Moreover, we may use \cite{Lyth}
\begin{eqnarray}\label{14}
\langle|\delta\psi|^2\rangle=\left(\frac{H}{2\pi}\right)^2.
\end{eqnarray} 
Using \eqref{x-def} and \eqref{dotgamma=0} we obtain
\be\label{be}
\dot{\psi}=\frac{\sqrt{\gamma}}{g}\dot{H}.
\ee 
Substituting \eqref{14} and \eqref{be} in \eqref{16}
together with the definition of $\varepsilon$ we obtain
\begin{eqnarray}\label{16'}
\Delta_R^2(k)=\left[\frac{g^2}{4\pi^2\gamma\varepsilon^2 }\right]|_{k=aH}=\left[\frac{H^2}{4\pi^2\psi^2\varepsilon^2 }\right]|_{k=aH},
\end{eqnarray}
where the last equality is achieved by using \eqref{x-def}. 
The scalar spectral index is obtained as
\begin{eqnarray}\label{18}
n_s=1+\frac{d\ln \Delta_{R}^2(k)}{d\ln k}&=&1-10\varepsilon+4\eta+2\delta.
\end{eqnarray}
Comparing with the recent Planck and BICEP2 measurements \cite{2-i,2-m,BI}
\begin{equation}\label{g(r)}
 \begin{array}{ll} n_s=0.9603\pm0.0073,\:\: 
\end{array}
\end{equation}
we obtain $\varepsilon\sim\eta < 10^{-2}$ and $\delta\ll10^{-2}$. The tensor power spectrum is also given by \cite{3}
\begin{eqnarray}\label{19}
\Delta^2_T=\left[\frac{2 H^2}{\pi^2}\right]|_{k=aH},
\end{eqnarray}
from which we obtain the spectral index $n_T$ as
\begin{eqnarray}\label{}
n_T=\frac{d\ln \Delta_{T}^2(k)}{d\ln k}=-2\epsilon.
\end{eqnarray}
The tensor to scalar ratio is given by
\begin{eqnarray}\label{21}
r=\frac{\Delta^2_T}{\Delta^2_R}=[8\varepsilon^2\psi^2]|_{k=aH}=\left[8\varepsilon^2\frac{\gamma
H^2}{g^2}\right]|_{k=aH}.
\end{eqnarray}
The recent Planck and BICEP2 measurements constrain the curvature perturbation power spectrum as \cite{2-i,2-m,BI}
 \begin{equation}\label{16''}
\Delta_R^2(k)\approx2.215 \times 10^{-9}.
\end{equation}
Approximating \eqref{16'} with this value we obtain
\begin{eqnarray}\label{16'''}
\left[\frac{g^2}{\gamma\varepsilon^2 }\right]|_{k=aH}\approx 8\times 10^{-8}.
\end{eqnarray}
Now, using \eqref{16'''} we find
\begin{eqnarray}\label{r}
r\approx H^2\times10^8.
\end{eqnarray}
Demanding $r$ to be consistent with the data probed by Planck ($r \leq 0.11$) and BICEP2 ($r \simeq 0.2 $) results respectively in $H^2\leq 11\times 10^{-10}$ and $H^2\approx 2\times 10^{-9}$. Moreover, requesting for $N_e\approx60$ with $\varepsilon\lesssim 10^{-2}$ in \eqref{Ne-gauge-flation} gives $\gamma\approx 6$. Substituting this $\gamma$ and $\varepsilon$ into \eqref{16'''} gives
\begin{eqnarray}\label{g}
g\approx 7\times10^{-6},
\end{eqnarray}
which has the same order of magnitude as in the {\it Chromo-natural} inflation.

\section{Conclusion}

In this paper, we have combined two recently developed ideas of {\it Gauge-flation} and {\it Kinetic coupled gravity}. Gauge-flation has been introduced to
establish inflation in the context of high energy physics where the slow-roll regime is set up by a non-Abelian gauge field $\textbf{A}$ with the field strength $\textbf{F}$. This class of models are based on a gauge field theory with $\textbf{F}^2$ and $\textbf{F}^4$ terms with a non-Abelian gauge group minimally coupled to gravity. Kinetic coupled gravity, on the other hand, has been introduced mainly to account for the current acceleration of the
universe. The non-minimal kinetic coupled gravity is set up by introducing the Einstein tensor besides the metric tensor within the $\textbf{F}^2$ term. By combining these ideas, we have presented a new class of inflationary models based on a non-Abelian gauge field theory with $\textbf{F}^2$ term non-minimally coupled to gravity. We have performed a perturbation analysis to confront the inflation under consideration with the recent Planck and BICEP2 results. We have found that the scalar spectral index is in agreement
with both Planck and BICEP2 results. Moreover, the tensor-to-scalar ratio is in agreement with the upper bound probed by Planck and by BICEP2 provided
respectively that $H^2\leq 11\times 10^{-10}$ and $H^2\approx 2\times 10^{-9}$.

\section*{Acknowledgment}
We would like to thank the anonymous referee whose useful comments 
improved very much the content of this paper. 
This research has been supported by Azarbaijan Shahid Madani university by a research fund No. 403.16.

\end{document}